\documentclass[aip,rsi,twocolumn,reprint,groupedaddress,amsmath,amssymb,a4paper]{revtex4-1}
\usepackage{graphicx}
\usepackage{dcolumn}
\usepackage{bm}
\usepackage{color}
\usepackage[separate-uncertainty = false]{siunitx} 
\DeclareSIUnit\ppm{ppm} 

\usepackage{float} 
\usepackage{verbatim} 

\newcommand{\Lapl}{\mathcal{L}}
\newcolumntype{.}{D{.}{.}{-1}} 
\begin{document}

\title{A straightforward 2$\omega$ technique for the measurement of the Thomson effect.}

\author{Isaac Ha\"{i}k Dunn}
\email{isaac.haik-dunn@ensicaen.fr}
\affiliation{Normandie Univ, ENSICAEN, UNICAEN, CNRS, CRISMAT, 14050 Caen, France}

\author{Ramzy Daou}
\email{ramzy.daou@ensicaen.fr}
\affiliation{Normandie Univ, ENSICAEN, UNICAEN, CNRS, CRISMAT, 14050 Caen, France}
 
\author{Colin Atkinson}
\affiliation{Schlumberger Gould Research Center, Cambridge, UK}
\affiliation{Department of Mathematics, Imperial College London, 180 Queen's Gate, London SW7 2AZ, UK}
\date{\today}

\begin{abstract}
We present a simplified, rapid, and accurate method for the measurement of the thermoelectric Thomson coefficient by the dynamical heating of a suspended wire by an alternating current. By applying a temperature gradient across the wire, we find that the response at the second harmonic of the excitation frequency is directly proportional to the Thomson coefficient. The absolute thermoelectric coefficient of a single material can therefore be extracted with high precision by a phase sensitive detector. We test our method on platinum and nickel wires and develop both analytical and numerical models to determine the leading sources of error.
\end{abstract}


\maketitle

\section*{Introduction}
The Thomson effect was hypothesized in the year 1851 and experimentally confirmed by William Thomson (Lord Kelvin) in 1853\cite{thomson18574,thomson1853}. It describes the generation or absorption of heat by charge carriers when a current is carried along a thermal gradient. It is one manifestation of the thermoelectric effect, which is also responsible for the Seebeck and Peltier effects.

The Thomson coefficient $\mu$ is related to the other thermoelectric coefficients by the Thomson relations $\mu = T dS/dT$ where $S$ is the Seebeck coefficient and $T$ is the absolute temperature, while the Peltier coefficient is $\Pi = ST$. These relations are the result of Onsager's reciprocity theorem \cite{onsager1931}. We would have little interest in the Thomson coefficient if it were not for the fact that it is the only thermoelectric coefficient that can be measured without reference to another material. It is therefore the only way to establish an absolute scale of thermoelectricity at temperatures above  $\sim$\SI{120}{\kelvin} where superconducting materials ($S = \mu = \SI[per-mode=symbol]{0}{\volt\per\kelvin}$) can be used as a reference. The Thomson heat should also be taken into account when evaluating the performance of thermoelectric devices\cite{teg-most-important}.

The particular challenge in obtaining $\mu$ experimentally is to separate the Thomson heat from the Joule heat, which arises whenever a current flows through a conductor. The Joule heat is proportional to the square of the current density, and is usually at least two orders of magnitude greater than the Thomson heat.

There has been little conceptual development in the measurement of the Thomson effect since the early work of Borelius \cite{borelius1928}. By reversing the current through a wire thermally clamped at both ends, he was able to detect the slight temperature changes due to the Thomson heat using a sensitive thermocouple. Similar apparatus was used by Lander\cite{lander1948} and Nystrom\cite{nystrom1947}. In this way an absolute scale of thermoelectricity was established up to \SI{2000}{\kelvin} \cite{cusack1958}.

Today, the modern standard was established by Roberts\cite{roberts1977absolute} who used a differential thermocouple to measure the temperature difference between two parallel insulated wires subject to the same temperature gradient, but with current flowing through them in opposite directions. This innovation improved the reliability of the experiment, as it was no longer necessary to subtract two large measurements from each other to obtain the small Thomson-related voltage.

Roberts' technique was applied to a number of materials including Pb, Pt, and Cu, over different temperature ranges. These experiments have formed the backbone of the absolute thermoelectric scale in the temperature range up to \SI{1400}{\kelvin}. For Pb in the range from \SIrange{0}{300}{\kelvin}, Roberts claimed an absolute accuracy of \SI[per-mode=symbol]{0.01}{\micro\volt\per\kelvin}.

Nevertheless, Roberts' experiment is both cumbersome and time consuming. In this article we present a new, simple technique that can be employed to measure the Thomson coefficient with relative ease and rapidity by heating a suspended wire using a low frequency alternating current. We validate the technique on high-purity nickel and platinum wires and compare our experimental results to approximate analytical solutions, more accurate finite element simulations, and to direct measurements of the Seebeck coefficient. 

Our calculations show that pure metals are the ideal subjects for the technique, as their low resistivity limits Joule heating effects. The measurements most closely follow analytical predictions when the resistivity is quite linear with temperature. We envisage that Thomson measurements on locally-available elemental metal wires could be used to produce calibration samples for Seebeck measurement apparatus. These might be used, for example, to improve on the specifications of commercial apparatus, or to calibrate a custom-built device.

\section*{Methods}
\sisetup{range-phrase=--}

\begin{figure}
\includegraphics[width=0.5\textwidth]{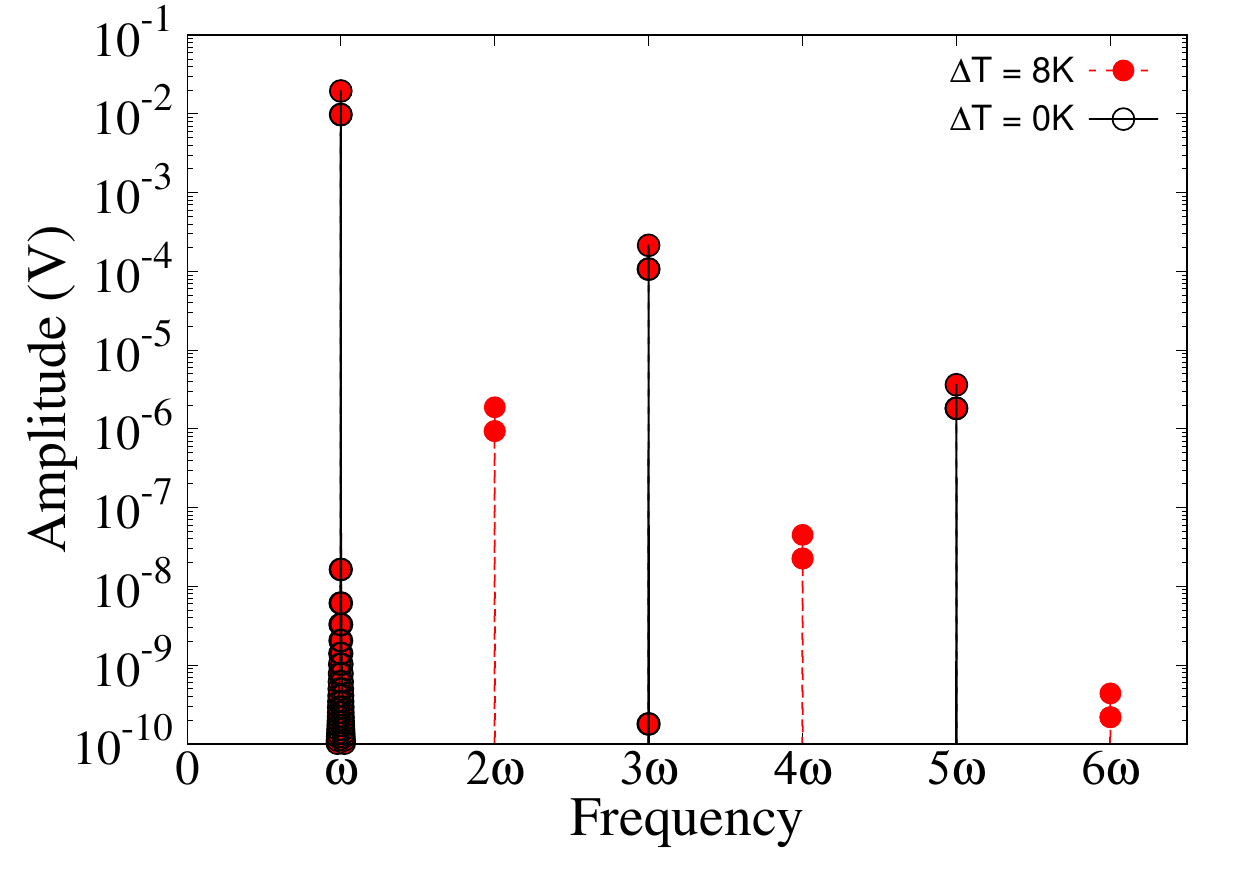}
\caption{Frequency spectrum of the magnitudes of voltage oscillations when the applied current is at a low angular frequency $\omega$, obtained from finite element simulations using a typical set of parameters. When $\Delta T = \SI{0}{\kelvin}$, there are signals at odd multiples of $\omega$ due to the resistance oscillations caused by Joule heating (black lines and open circles). When $\Delta T = 8$K, there are additional peaks at even multiples of $\omega$ arising from the Thomson heating, the largest of which is at $2\omega$ (red dashed lines and filled circles). The odd-multiple signals are still present and are unaffected by the Thomson effect.}
\label{fig:fft}
\end{figure}

The heat equation of a homogeneous material through which both heat and electric current flow is:
\begin{equation}
\nabla \cdot (\kappa \nabla T) - \mu \vec J \cdot \nabla T +\rho J^2 = C_V \frac{\partial T}{\partial t}
\label{eqn:fullheat}
\end{equation}
where $\kappa$ is the thermal conductivity, $T$ is the temperature, $\mu$ is the Thomson coefficient, $\vec{J}$ is the current density, $\rho$ is the resistivity, $C_V$ is the heat capacity per unit volume and $t$ is time. The second and third terms are the Thomson and Joule heat generated per unit volume, respectively. 

We are interested in the particular case of a suspended wire of length $2l$ and cross-section $A$ thermally clamped at both ends carrying a sinusoidally driven current density, $J=J_1\sin\omega t$. The wire diameter is much less than $2l$, which permits a one-dimensional treatment. The boundary conditions are such that there is a static thermal gradient imposed along the length of the wire, i.e. $T(-l)=T_0-\frac{\Delta T}{2}$ and $T(l)=T_0+\frac{\Delta T}{2}$.

In this configuration, we can use the resistance of the wire as measured between the thermally clamped ends as a probe of the self-heating. This principle has been exploited before to measure the thermal conductivity of thin wires and substrates in what is known as the ``$3\omega$ technique''\cite{Cahill1990,dames20051} (where $\Delta T=\SI{0}{\kelvin}$), but has not until now been used to determine the Thomson coefficient.

If the resistivity of the wire depends on temperature, then under these boundary conditions the resistance of the wire oscillates due to the oscillating Joule and Thomson heating. The Joule heating is always positive and leads to resistance oscillations at frequency $2\omega$, since $\rho J^2 = \rho\frac{J_1^2}{2}(1-\cos 2\omega t)$.

There are then two contributions to the Thomson heat. The first arises from the oscillating temperature profile generated by the Joule heating. Since this profile is symmetric about $x=0$, $dT/dx$ changes sign at $x=0$ and so the Thomson heat is asymmetric about $x=0$. Hence the resistance oscillations induced by this term (which would be at frequencies $\omega$ and $3\omega$) are effectively canceled when integrated along the length of the wire. The second contribution arises from the applied static temperature gradient, $dT/dx = \Delta T/2l$. Since this gradient is asymmetric about $x=0$, finite resistance oscillations occur at a frequency $\omega$.

The above description is not exact; temperature oscillations at higher multiples of $\omega$ also occur as the Joule and Thomson heat respond to the oscillatory changes in resistance.

All of the resistance oscillations are probed by the same current that causes the heating, leading to voltage signals at frequencies $3\omega$ and $2\omega$ arising primarily from the Joule and Thomson heating respectively. The frequency spectrum with and without the applied gradient $\Delta T$ is shown in Figure~\ref{fig:fft}. In the low frequency limit, the amplitude of the $3\omega$ voltage is determined almost exclusively by $\kappa$, and by $\mu/\kappa$ for the $2\omega$ voltage. 

\begin{figure}
    \includegraphics[width=0.47\textwidth]{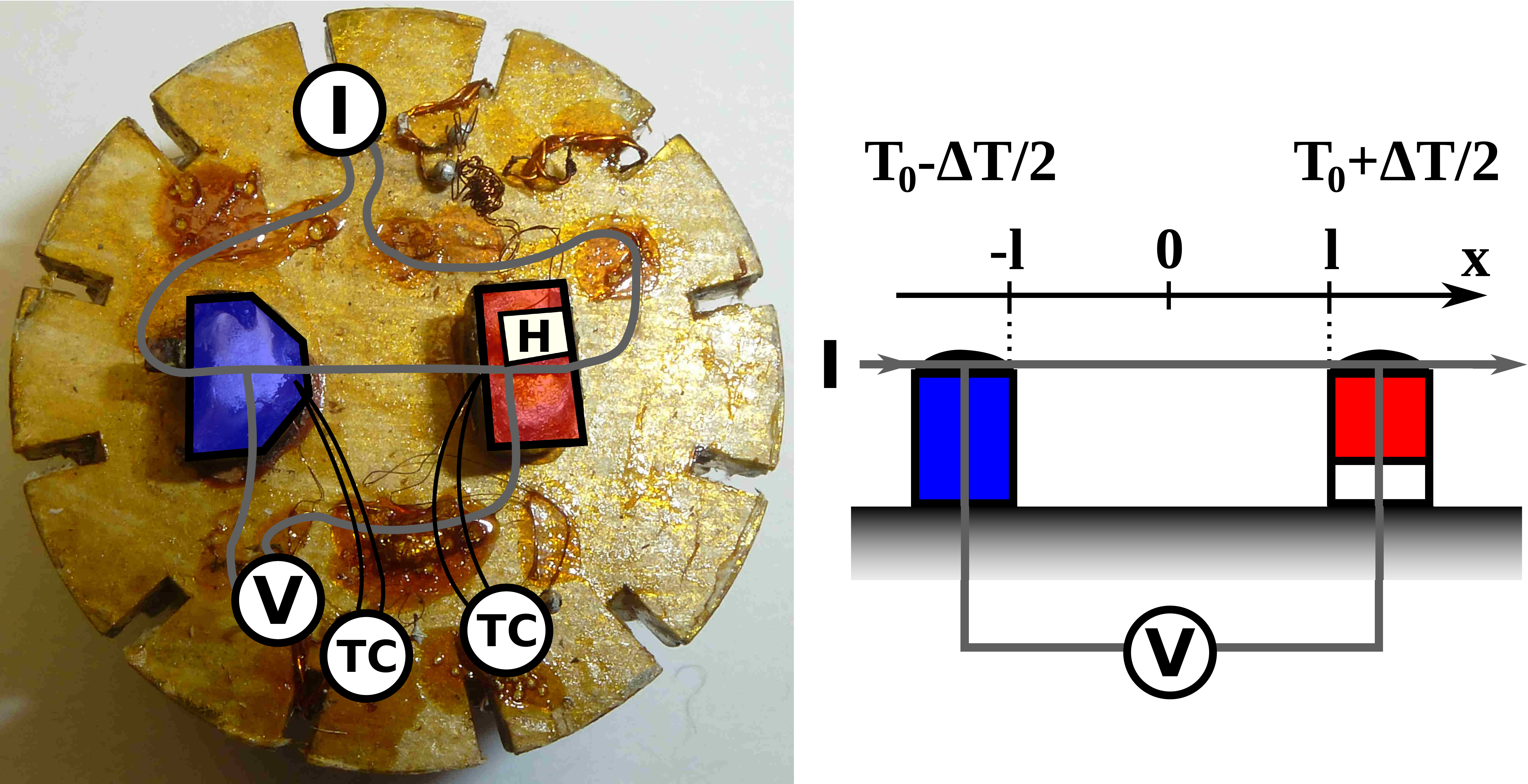}    
\caption{Experimental configuration. Left: Photograph of the PPMS sample holder overlaid with schematic of critical components. The diameter is \SI{27}{\milli\meter}. Right: side-view schematic. The \SI{25}{\micro\meter}-diameter wire under test (Ni or Pt, grey line) runs continuously between the current contacts (I), which are well thermalized to the cryostat temperature. The central part of the wire (of length $2l=5$\si{\milli\meter}) is suspended between two temperature-controlled blocks. Additional pieces of the same wire are used to link the wires crossing the blocks to the isothermal voltage contacts (V). The hot side temperature is controlled using a feedback loop between the temperature measured by the thermocouple (TC) and the resistive heater (H). See text for further details.}
\label{fig: puck_setup}
\end{figure}

To be more quantitative, in the case that the resistivity can be approximated by a linear function near the temperature of interest, $\rho = \rho_0 + \rho_1 T$, and if $\kappa$ and $\mu$ do not depend on temperature, we obtain the steady-state solution of Equation \ref{eqn:fullheat} (see Appendix \ref{sec:solutions}). The resulting voltage across the wire can be expanded to leading order in $\mu$ to give:
\begin{equation}
 \mu = -\frac{RI}{2\Delta T}\lim_{\omega\to 0} \left(\frac{V_{2\omega}}{V_{3\omega}}\right) \left[ 1 - \frac{\xi^2}{60} - \frac{\xi^4}{4200} + O(\xi^6) \right]
 \label{eqn:approximation}
\end{equation}
where $\xi^2 = \alpha I^2 R/K$, $R$ is the resistance of the wire, $K = \kappa A/2l$ is its thermal conductance, $\alpha = 1/R(dR/dT) = \rho_1/\rho$ is the temperature coefficient of resistance and $I$ is the applied current. All values are taken at $T_0$.

We find that the corrections are small, typically no more than $\SIrange[range-phrase=-]{2}{3}{\percent}$ in the regime where we operate. The Thomson coefficient can therefore be determined from measurement of the two voltages $V_{2\omega}$ and $V_{3\omega}$ in the limit of low frequency. The corrections can be evaluated if $K$ is extracted from $V_{3\omega}$, which depends only weakly on $\mu$. 
The dimensions of the wire are not required to obtain $\mu$, as has been observed before \cite{amagai2015ac}.
The only preliminary measurement necessary is the resistance of the wire as a function of temperature.

In practice, we use a finite element method incorporating the measured temperature-dependent resistance of the wire to more precisely simulate the temperature profiles and thus the voltages generated. 

For comparison, we measured the Seebeck coefficient of the wires directly by a standard steady-state method involving two type-E thermocouples. The reference leads for the thermoelectric voltage are the chromel legs of the thermocouples. The absolute Seebeck coefficient of chromel is compiled from various sources and subtracted\cite{Chiang1974,Chaussy1981}.

The experimental configuration is shown in Figure~\ref{fig: puck_setup}. The wire under investigation was mounted between two copper blocks. A continuous length was used between the isothermal current contacts, while additional pieces were used to connect the isothermal voltage leads in a four-point probe configuration. The current and voltage wires met on top of two copper blocks, where they were soldered in place. The cold side block (a copper cube $\sim$\SI{3}{\milli\metre} across) was glued using GE7031 varnish to the sample holder (or puck) of a Physical Properties Measurement Systems (PPMS) cryostat. The surface of the PPMS puck was electrically isolated everywhere using a slip of cigarette paper soaked in varnish. The hot side block stood on a square of glass \SI{1}{\milli\metre} thick. The blocks are $\sim$\SI{5}{\milli\metre} apart. The free ends of the wires were glued to the surface of the PPMS puck and then soldered to \SI{0.2}{\milli\metre} copper wire contacts. In this way we ensured that there is no contribution to the voltage measured in the wire from the Seebeck effect and that the suspended part of the wire was thermally well clamped. The hot block was heated by a small RuO$_2$ chip resistor and the temperature of the block was measured using a type-E fine wire thermocouple. The sample holder was top-loaded into the PPMS chamber, which was then pumped to high vacuum.

The resistance of the suspended wire was measured using a lockin amplifier in the range \SIrange{10}{400}{\kelvin} using a small (\SI{100}{\micro\ampere}) alternating current to avoid self-heating. This served as a calibration curve. The measured value reflects almost exclusively the resistance of the part of the wire freely suspended between the blocks. The massive solder and copper blocks have such low resistance that very little signal arises from the parts of the wire within the solder.

Alternating currents of up to \SI{100}{\milli\ampere} were then applied to the suspended wire using a Keithley 6221 current source. The resulting voltage across the wire was measured using a lockin amplifier at \numlist{1;2;3}$\omega$. A DC nanovoltmeter read the temperature of the hot side thermocouple and a software control loop adjusted the DC current applied to the chip heater to ensure stability to better than \SI{\pm 5}{\milli\kelvin}. Experiments were performed at \SI{5}{\kelvin} or \SI{10}{\kelvin} intervals in the range \SIrange{25}{400}{\kelvin}.

Our measurement protocol first took \num{13} logarithmically spaced points in the range \SIrange{0.1}{10}{\hertz} at three different currents, focusing on the response at $3\omega$ in order to estimate the temperature dependence of the thermal conductivity using a linear model, $\kappa = \kappa_0+\kappa_1(T-T_0)$, where $\kappa_1 = \frac{d\kappa}{dT}|_{T_0}$. A small temperature difference ($\Delta T = \SI{2}{\kelvin}$) was applied. Exemplary data is shown in Figure~\ref{fig:fulldata}d.

Then at the highest current, the $2\omega$ voltages were measured for $\Delta T$ = \SIlist{2;6;10}{\kelvin} over the same frequency range (Figs.~\ref{fig:fulldata}a,~\ref{fig:fulldata}b).

We confirmed that the $3\omega$ voltages do not depend on $\Delta T$ as long as the average temperature is kept constant, e.g. for $T_0 = \SI{300}{\kelvin}$, the PPMS (cold side) temperatures were \SI{299}{\kelvin}, \SI{297}{\kelvin} and \SI{295}{\kelvin}. To save time, we did not systematically acquire $3\omega$ voltages for $\Delta T$ = \SIlist{6;10}{\kelvin}.

Since analytical solutions of Equation~\ref{eqn:fullheat} are limited to special cases, a finite-element, finite-difference numerical simulation of the experiment was developed in Python using the FEniCS package \cite{ans20553}. This was used to extract material parameters by least-squares fitting of the simulated results to the experimental data, and to test for the impact of various sources of error. The simulations are described in more detail in Appendix \ref{sec:finite}. The greatest departure from the analytical solution comes from non-linearity of the resistivity of the wire, which is much more significant in nickel than in platinum. The second most important correction arises from the relatively large heating applied, which leads to significant temperature rises in the wire (on the order of \SI{10}{\kelvin} for the highest currents and temperatures). In this regime the temperature dependence of the thermal conductivity starts to play a significant role and must be included in the model.

We simulated the effect of a small finite DC offset current, as is expected from the current source employed. Simulations and experimental results showed that a signal similar in magnitude to the Thomson signal is expected in the $2\omega$ voltage. However, this signal does not depend on $\Delta T$ and can therefore be eliminated by taking differences such as $V_{2\omega}(\Delta T = \SI{10}{K}) - V_{2\omega}(\Delta T = \SI{2}{K})$. These differences are shown in Figure~\ref{fig:fulldata}c for the experimental data. The frequency dependence of these curves now resembles the thermal response seen in the $3\omega$ data, as expected.

We also studied the effect of thermal radiation losses and found them to be insignificant over this temperature range. In the Appendix \ref{subsec: radlosses} we predict an upper temperature limit to the utility of the experiment based on the simulations.

\section*{Results and Discussion}

\begin{figure}[!htbp]
\includegraphics[width=0.48\textwidth]{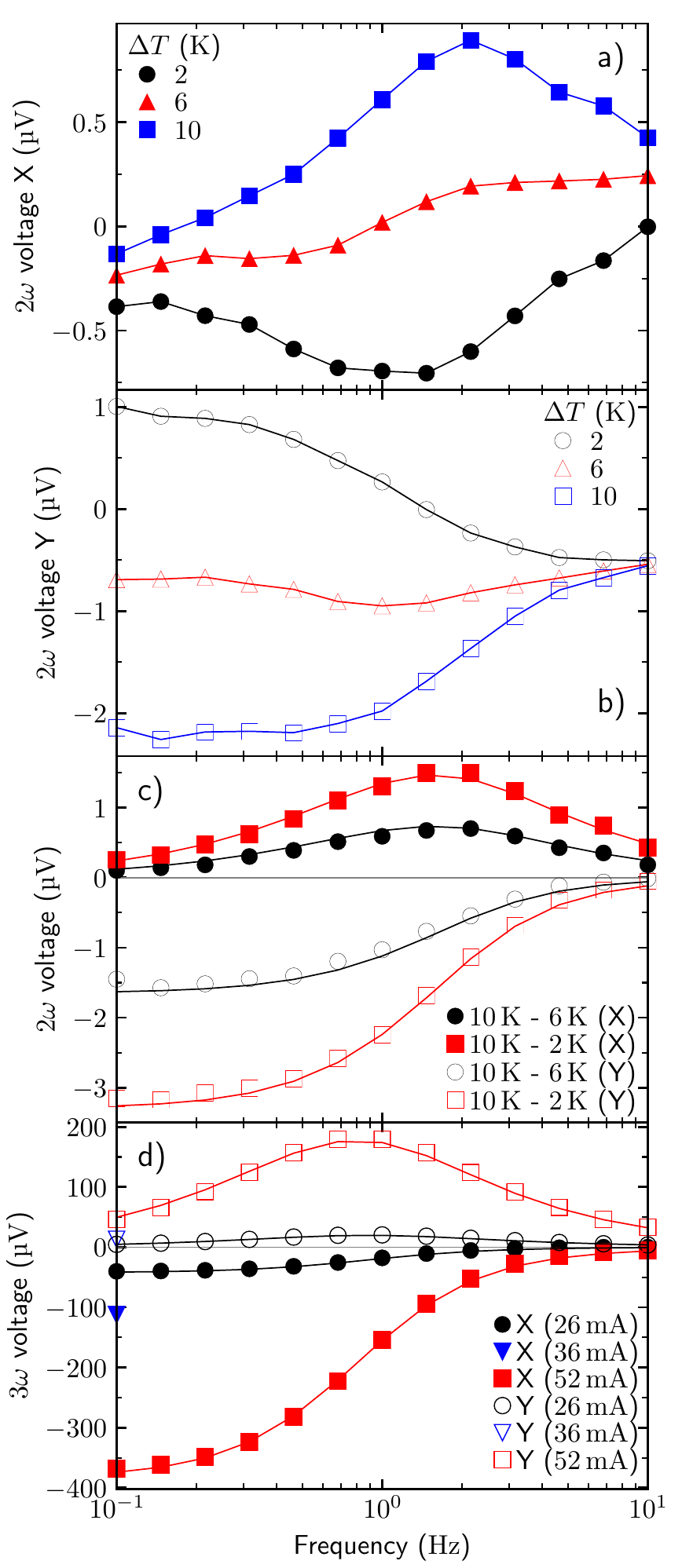}
\caption{A typical data set for Ni at $\SI{170}{\kelvin}$.
a) and b): In-phase (X) and out-of-phase (Y) parts of raw $V_{\num{2}\omega}$ data as a function of frequency, taken for $\Delta T =$ \SIlist{2;6;10}{\kelvin} with $I_\text{ac} =\SI{52}{\milli\ampere}$ and $I_\text{dc} =\SI{80}{\micro\ampere}$.
c): The difference of the $2\omega$ voltages at different values of $\Delta T$ leaves only the Thomson signal and eliminates the signal due to the residual DC offset of the current source. The solid lines are fitted curves obtained by simulations. d) The raw $3\omega$ voltages (X and Y components) taken for three different values of $I_\text{ac}$.}
\label{fig:fulldata}
\end{figure}

At every temperature point, we acquired data sets such as the one presented in Fig.~\ref{fig:fulldata} and analysed them.  The raw data shows that it is straightforward to obtain the small $2\omega$ signals ($\sim\SI{1}{\micro\volt}$) with good resolution even in the presence of the large $1\omega$ ($\sim\SI{10}{\milli\volt}$) and $3\omega$ ($\sim\SI{100}{\micro\volt}$) signals. This clearly shows the advantage of the spectral separation of the different signals and the role of phase sensitive detection.

\begin{figure}[!htbp]
\includegraphics[width=0.48\textwidth]{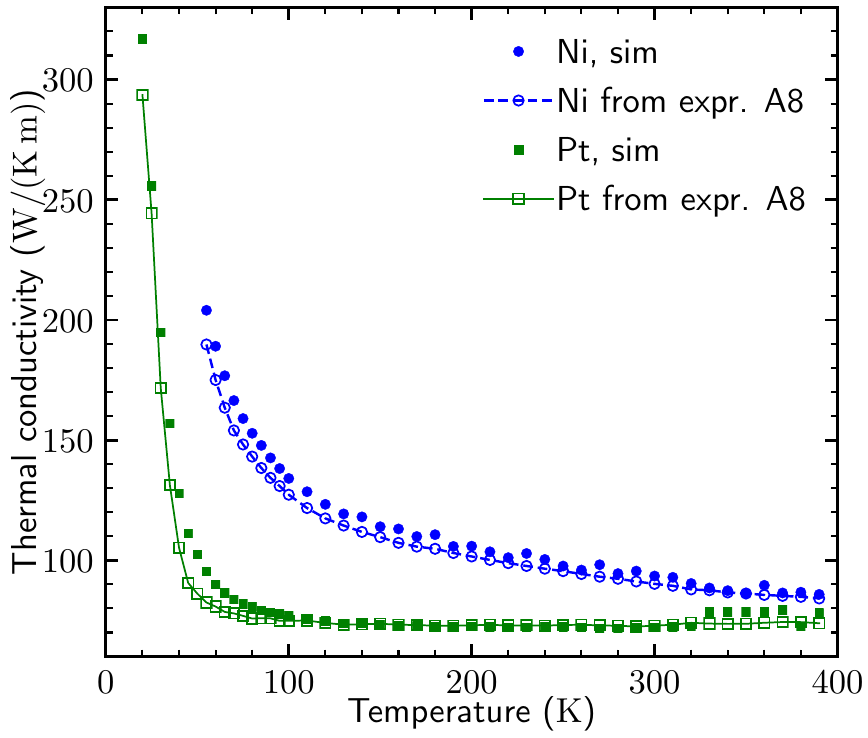}
\label{kappa_vs_T}
\caption{$\kappa$ extracted from $3\omega$ data using finite element simulations, as well as via the three leading terms of Eqn.~\ref{eqn: correction_thermal_capacitance}. The agreement is best when the resistivity is linear in temperature and the thermal conductivity is a constant. These conditions are most closely followed for Pt in the range \SI{150}{\kelvin}-\SI{400}{\kelvin}.}
\label{fig:kappavsT}
\end{figure}

The first step of our analysis is always to fit the $3\omega$ data to obtain the thermal conductance $K$. By assuming that the literature value of resistivity\cite{crchandbook} is correct, we obtain a sample geometric factor $A/2l$ which allows us to obtain the thermal conductivity, $\kappa$ from $K = \kappa A/2l$. The values of $\kappa$ for the Ni and Pt wires are shown in Figure~\ref{fig:kappavsT}, evaluated by fitting the data using the finite element simulations, and also compared to the values given by Eqn.~\ref{eqn: correction_thermal_capacitance}. The values of $\kappa$ obtained are close to the tabulated values\cite{crchandbook}. These are compared directly at \SI{300}{\kelvin} in Table~\ref{tab:summary}. It is worth noting here that the literature values of resistivity for both Ni and Pt are given with an uncertainty of around 3\%, while the values of $\kappa$ have an uncertainty of around $6\%$. Our derived values are well within these bounds. We reiterate at this point that the absolute value of $\kappa$ is not required to obtain the Thomson coefficient, it is presented here as a check on the general validity on the experiment.

\sisetup{group-digits=integer}
\begin{table}
\begin{tabular}{l | . . }
Value at \SI{300}{\kelvin} & \text{Ni} & \text{Pt}  \\
 \hline
 \hline
R (\si{\ohm}) & 0.8209(1) & 1.0185(1) \\
I (\si{\milli\ampere})\footnotemark[1]  & 30.0 & 35.2 \\
$\rho$ (\si{\micro\ohm\centi\meter})\footnotemark[2] & 7.20 & 10.8 \\
$A/2l$ ($10^{-8}$\si{\meter}) & 8.771(1) & 10.604(1) \\
$\lim_{\omega \rightarrow 0}Y_{2\omega}$ (\si{\micro\volt})\footnotemark[3] & -0.33(2) & -0.23(3)\\
$\lim_{\omega \rightarrow 0}X_{3\omega}$ (\si{\micro\volt}) & -265(1) & -457(1)\\
\hline
$\kappa$ (\si[per-mode=symbol]{\watt\per\kelvin\per\meter})\footnotemark[2] & 91.7 & 71.6 \\
$\kappa$ ($J_1\rightarrow 0$)\footnotemark[1]$^,$\footnotemark[4] & 85.8(2) & 68.6(3) \\
$\kappa$ ($\omega \rightarrow 0$)\footnotemark[5] & 90.2(2) & 72.8(3) \\
 $\kappa$ (simulations) & 93.5(5) & 72.2(5) \\
 \hline
 $\mu$ (\si[per-mode=symbol]{\micro\volt\per\kelvin})\footnotemark[6] &  & -9.1\\
 $\mu$ ($J_1\rightarrow 0$)\footnotemark[7] & -15.6(4) & -9.0(3)\\
 $\mu$ ($\omega \rightarrow 0$)\footnotemark[8] & -15.5(4) & -9.0(3)\\
 $\mu$ (simulations) & -15.4(4) & -8.5(4)\\
 \hline
 $S$ (\si[per-mode=symbol]{\micro\volt\per\kelvin})\footnotemark[6] & & -4.92(1) \\
 $S_0 + \int \mu/T dT$ (\si[per-mode=symbol]{\micro\volt\per\kelvin}) & -18.49(8) & -4.67(4) \\
 $S_\text{meas}$ (\si[per-mode=symbol]{\micro\volt\per\kelvin}) & -18.7(1) & -4.45(2) \\
 \hline
\end{tabular}
\footnotetext[1]{Highest of the three currents used.}
\footnotetext[2]{Ref.~\onlinecite{crchandbook}}
\footnotetext[3]{Av. values for $\Delta T=\SI{1}{\kelvin}$}
\footnotetext[4]{Eqn.~\ref{eqn:kappalumped}}
\footnotetext[5]{Eqn.~\ref{eqn: correction_thermal_capacitance}}
\footnotetext[6]{Ref.~\onlinecite{roberts1981}}
\footnotetext[7]{Eqn.~\ref{eqn:mulumped}}
\footnotetext[8]{Eqn.~\ref{eqn:approximation}}
\caption{Experimental and literature values at \SI{300}{\kelvin}. The resistance of the wire is known with the least uncertainty. The literature value of $\rho$ is used to determine the geometric factor $A/2l$, which is required to obtain $\kappa$, but which is not required to extract $\mu$. The values of $Y_{2\omega}$ and $X_{3\omega}$ in the low frequency limit can be used to obtain $\kappa$ and $\mu$ depending on which approximations are made (see text and appendices for details). The values obtained from numerical simulations incorporate the fewest approximations and are taken to be the most definitive. The Thomson coefficient $\mu$ is integrated to obtain the Seebeck coefficient via the Kelvin relation and compared to the Seebeck coefficient $S_\text{meas}$, measured directly against reference leads.}
\label{tab:summary}
\end{table}

\begin{figure}[!htbp]
\includegraphics[width=0.48\textwidth]{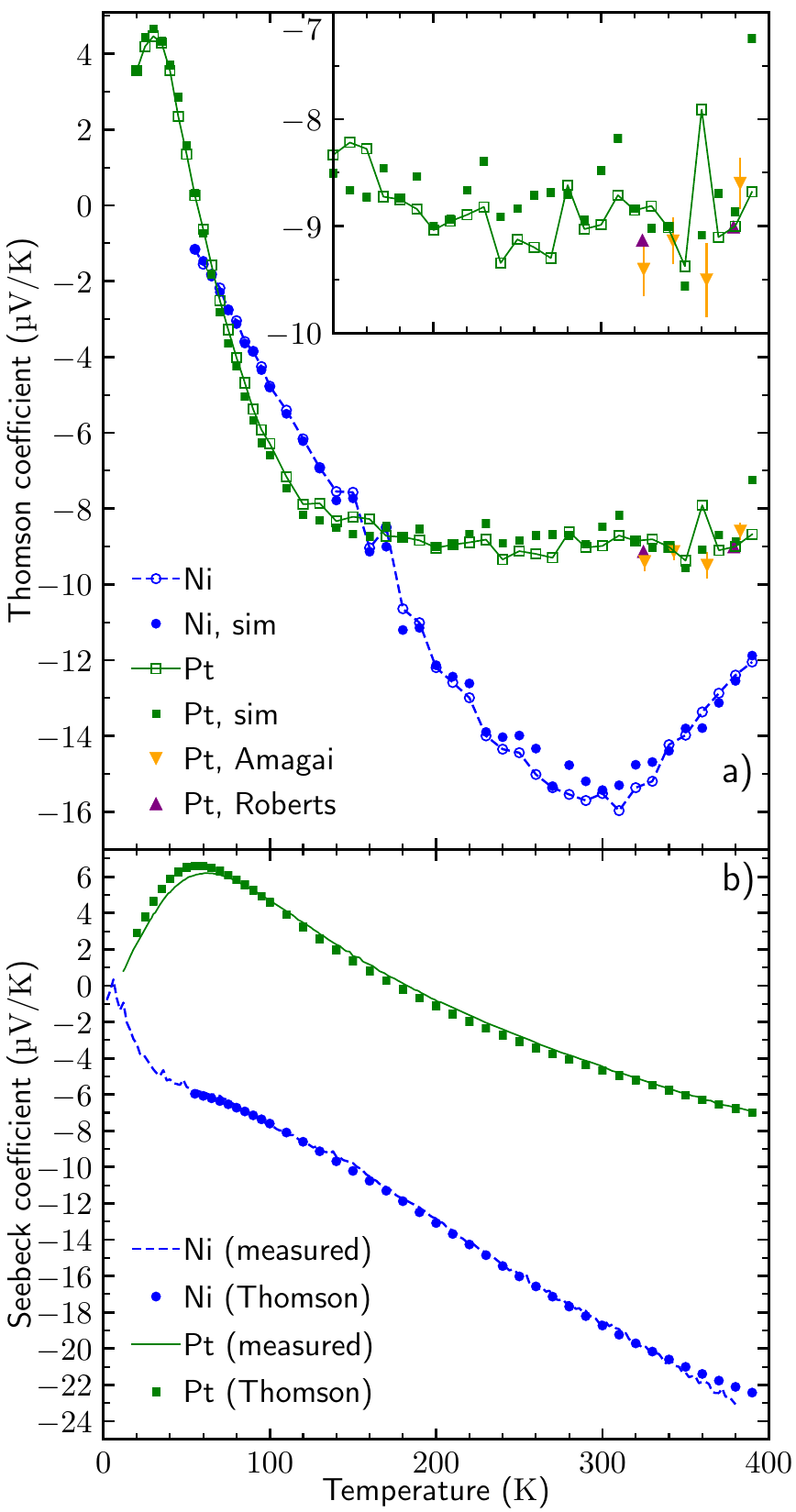}
\caption{a) Thomson coefficient of nickel and platinum as a function of temperature, obtained by fitting numerical simulations to the data and from Eqn.~\ref{eqn:approximation}. The agreement between the two methods to obtain $\mu$ is very good, suggesting that the analysis leading to expression Eqn.~\ref{eqn:approximation} is sound. Inset: enlargement of the high temperature data for Pt, compared to points from Refs.\onlinecite{roberts1977absolute,amagai2015ac}.
b) Seebeck coefficients of nickel and platinum as a function of temperature. Direct measurements using reference leads (whose contribution has been subtracted) are compared to S derived from the Thomson data in a) via the Kelvin relation. The offset in $S(T)$ arising because $\mu$ was not measured down to $\SI{0}{\kelvin}$ was chosen such that the least square error between the $S(T)$ curves would be minimized.}
\label{fig: mu_vs_T_Ni}
\end{figure}

The next step in the analysis is to fit the $2\omega$ voltages to obtain values of the Thomson coefficient $\mu$ using the finite element simulations. The values of $\mu$ shown in Figure~\ref{fig: mu_vs_T_Ni}a and are compared to the values given by Eqn.~\ref{eqn:approximation}. The sample dimensions do not play a role. The difference between the two methods of obtaining $\mu$ is smaller where the resistivity is more linear with temperature; and where the thermal conductivity is less temperature-dependent. These are the conditions where the approximations involved in obtaining Eqn.~\ref{eqn:approximation} are more closely followed. For both $\mu$ and $\kappa$ we consider the values fitted to finite element simulations to be the most definitive. 

In Table~\ref{tab:summary} we also compare the values of $\mu$ and $\kappa$ extracted using the approximate solutions to the heat equation derived in the appendices. For the small current limit $J_1 \rightarrow 0$ (equivalent to the results shown in Refs.~\onlinecite{dames20051,Lu_lumped_2001}), the values of $\kappa$ deviate very strongly. This is because we are working in the regime where the current is relatively large, and the self-heating of the wire can no longer be considered homogeneous. When a temperature-dependent resistivity is incorporated into the heat equation (Appendix~\ref{sec:solutions}.1), the analytical solution in the limit $\omega \rightarrow 0$ provides a considerable improvement, within 2-3\% of the finite element results which we take to be definitive.

To obtain the Seebeck coefficients from $\mu$ via the Kelvin relation, we perform the numerical integration $S(T) = \int_{T_\text{ref}}^T \mu(T^\prime)/T^\prime dT^\prime + S(T_\text{ref})$ where $S(T_\text{ref})$ is the value of the Seebeck coefficient at temperature $T_\text{ref}$. These are shown in Figure~\ref{fig: mu_vs_T_Ni}b, and compared with the direct measurements of $S$ against a known reference material.

Agreement between the values of $\mu$ obtained here and the values from Refs.~\onlinecite{roberts1977absolute,amagai2015ac} is quite good. When integrating to obtain the Seebeck coefficient, much of the random noise is suppressed and the temperature dependence of the curves match well over a large range. 

\section*{Conclusions}
We have presented a new phase-sensitive technique to obtain the Thomson coefficient to within $\pm 0.4$ \si[per-mode=symbol]{\micro\volt\per\kelvin} over the temperature range \SIrange{20}{400}{\kelvin}. We demonstrated the technique on nickel and platinum wires and compared our values to the results that have been used as a standard in recent times\cite{roberts1977absolute} as well as more recent efforts\cite{amagai2015ac}. Phase-sensitive detection has distinct advantages over DC measurement when small signals must be extracted from a large background, the most important of which is the separation of small and large signals in the spectral response. In Appendix~\ref{sec:dc} we discuss the challenges facing a DC implementation of this technique. 

The experiment is simple and compact enough that it could be implemented in any reasonably equipped laboratory to provide a local calibration of the thermoelectric scale sufficiently accurate for most requirements. In particular it does not require the use of a direct temperature probe that might disturb the temperature profile in the wire or otherwise introduce inaccuracy. Such a technique was recently implemented in Ref.~\onlinecite{amagai2015ac}, which we review briefly and compare to our work in Appendix~\ref{sec:amagai}.

While we have relied on extensive modelling of the data using finite element simulations, we have also shown that we can determine the Thomson coefficient with reasonable accuracy using several analytical approximations. If accuracy is not paramount, only very limited computational resources are required.

The analytical solutions do serve to give us some insight as to the feasibility of measuring other materials. While it would be of great practical interest to measure directly a common thermoelectric reference lead such as chromel, the temperature coefficient of resistance is ten times smaller than that of platinum, while the resistivity is ten times greater. This means that only a small current could be used in order to limit the temperature rise to a typical \SI{10}{\kelvin} at \SI{300}{\kelvin}, leading to very small signals. We estimate that for the same geometry as the wires presented here, the Thomson signal for chromel would be no greater than \SI{50}{\nano\volt}. Similar constraints likely apply to most alloys. A semiconductor with a strong negative temperature coefficient of $\rho$ and a large Thomson coefficient might be possible to measure, but such materials are often difficult to obtain as wires which can be easily contacted electrically and thermally. While the method may be of limited scope, it is relatively straightforward to implement on common materials.

The technique should also be feasible at higher temperatures. We estimate that at $\sim$\SI{800}{\kelvin} the error in $\mu$ arising from thermal radiation losses will be of order 1\%, within the current level of noise. The miniature nature of the experiment helps to limit this effect.

The accuracy could no doubt be improved to be competitive with the established absolute scale \cite{roberts1977absolute}. Longer acquisition times could be used to reduce noise, particularly at low frequencies. Thermometers calibrated with reference to a primary standard could be used at both temperature controlled ends of the suspended wire, instead of the thermocouples that we have used for simplicity. A resistance standard could also be employed to improve the accuracy of the preliminary resistance measurement. These measures should be straightforward at a standards laboratory.

\appendix
\section{Solutions to the heat equation}
\label{sec:solutions}

No closed form solution for Equation~\ref{eqn:fullheat} has been found. 
A full time-dependent solution to Equation \ref{eqn:fullheat} under sinusoidal excitation is not strictly necessary to establish $\mu$ or $\kappa$. Practically, however, some understanding of the frequency dependence is useful in order to extrapolate the spectrally separated voltages to the DC limit.
In this appendix we present the various exact and approximate solutions that we have developed to treat the problem, and relate these to previous work.

\subsection{Exact steady-state solution}
In order to relate the oscillatory voltages in the DC limit to $\mu$ and $\kappa$, we solve Equation~\ref{eqn:fullheat} in the case where the right hand side is zero, $\mu J_1 \sin(\omega t) \rightarrow \mu J_1$ and $\rho=\rho_0+\rho_1 T$.
In this case, the temperature profile along the wire is given by:
\begin{equation}
\begin{split}
 T&(x) = e^{\lambda x}\Bigg \{ \frac{\cos\psi x}{\cos\psi l}
 \left[ \bigg( T_0+\frac{\rho_0}{\rho_1} \bigg )\cosh\lambda l -\frac{\Delta T}{2}\sinh\lambda l \right] \\
 & + \frac{\sin\psi x}{\sin\psi l}
 \bigg [ -\bigg( T_0+\frac{\rho_0}{\rho_1} \bigg )\sinh\lambda l +\frac{\Delta T}{2}\cosh\lambda l \bigg ] 
 \Bigg \} - \frac{\rho_0}{\rho_1}
\end{split}
\end{equation}
where $\lambda=\mu J/2\kappa$ and $\psi = \lambda\sqrt{4\kappa\rho_1/\mu^2 -1}$. The resulting voltage across the wire is:
\begin{equation}
 \begin{split}
  V(J) &= \int_{-l}^{l} \rho(T) J dx = 2l\rho_0 J + \rho_1 J\int_{-l}^{l} T(x) dx\\
    &= \frac{2\kappa}{J} \Bigg \{ 
\frac{\Delta T}{2} \left[ \lambda - \psi\frac{\sinh(2l\lambda)}{\sin(2l\psi)} \right]\\
&+\left( T_0 +\frac{\rho_0}{\rho_1} \right) \psi \left[ \frac{\cosh(2l\lambda)-\cos(2l\psi)}{\sin(2l\psi)} \right] \Bigg \}
 \end{split}
\end{equation}
Note that $(T_0+\rho_0/\rho_1) = \rho/\rho_1$, and $\rho_1 = \rho\alpha$ where $\alpha = \frac{1}{R} \frac{dR}{dT}$ is the temperature coefficient of resistance evaluated at $T_0$. The voltage can then be rewritten in terms of quantities that do not depend on the wire dimensions:
\begin{equation}
 \begin{split}
  V(I) &= \frac{2K}{I} \Bigg [ \frac{\Delta T}{2} \bigg (\Lambda - \Psi\frac{\sinh\Lambda}{\sin\Psi} \bigg )\\
  & + \frac{1}{\alpha} \Psi \bigg( \frac{\cosh\Lambda-\cos\Psi}{\sin\Psi} \bigg ) \Bigg ]
 \end{split}
 \label{eqn:nondimvoltage}
\end{equation}
where $\Lambda = \mu I/2K$ and $\Psi = \Lambda\sqrt{4KR\alpha/\mu^2 -1}$.

In this formulation, it is not immediately obvious which terms would correspond to the spectrally separated voltage signals at $\omega$, $2\omega$ and $3\omega$ if Eqn.~\ref{eqn:fullheat} were taken to the low frequency limit. We exploit the symmetry of the Thomson effect to isolate the signal that arises from the Thomson heat to make the following identification:
\begin{equation}
 \begin{split}
  \lim_{\omega\to 0} V_{2\omega}(I) &\approx \frac{V(I)+V(-I)}{4} \\
  &= \frac{\mu\Delta T}{4}\left[ 1 - \bigg( \frac{4KR\alpha}{\mu^2} - 1\bigg)^{1/2}\frac{\sinh\Lambda}{\sin\Psi} \right]
 \end{split}
\end{equation}
The remaining components change sign when the current is reversed. This is the behaviour expected for the undisturbed resistance as well as the additional resistance generated by Joule heating. The resistance of the undisturbed wire can be subtracted from this to leave the part of the resistance that would oscillate at $2\omega$. The amplitude of oscillation would be half of this value, and this amplitude is further halved in the $3\omega$ voltage when it is mixed with the sinusoidal current. Therefore:
\begin{equation}
 \begin{split}
  \lim_{\omega\to 0} &V_{3\omega}(I) \approx \frac{V(I)-V(-I)}{8} - \frac{IR}{4} \\
  &= \frac{\mu}{4\alpha}\left( \frac{4KR\alpha}{\mu^2} - 1 \right)^{1/2} \left( \frac{\cosh\Lambda-\cos\Psi}{\sin\Psi} \right) -\frac{IR}{4}
 \end{split}
 \label{eqn:V3approx}
\end{equation}

If we now take the ratio of these two voltages and perform a series expansion around $\mu$, we find:
\begin{equation}
 \lim_{\omega\to 0}\frac{ V_{2\omega}(I)}{ V_{3\omega}(I)} \approx \frac{\mu\Delta T}{RI} 
 \left[ \frac{\xi\sin\xi -\xi^2}{2(1-\cos\xi) -\xi\sin\xi} \right] + O(\mu^3)
\end{equation}
where $\xi = I \sqrt{R\alpha/K}$. When terms in $O(\mu^3)$ and higher are neglected this can be rearranged to give:
\begin{equation}
 \mu \approx \frac{RI}{2\Delta T} 
 \lim_{\omega\to 0} \frac{ V_{2\omega}(I)}{ V_{3\omega}(I)} 
 \left[ \frac{2(1-\cos\xi) -\xi\sin\xi}{\xi\sin\xi -\xi^2} \right]
\end{equation}
where a series expansion with respect to $\xi$ leads directly to Equation \ref{eqn:approximation}.

\begin{figure}[H]
\includegraphics{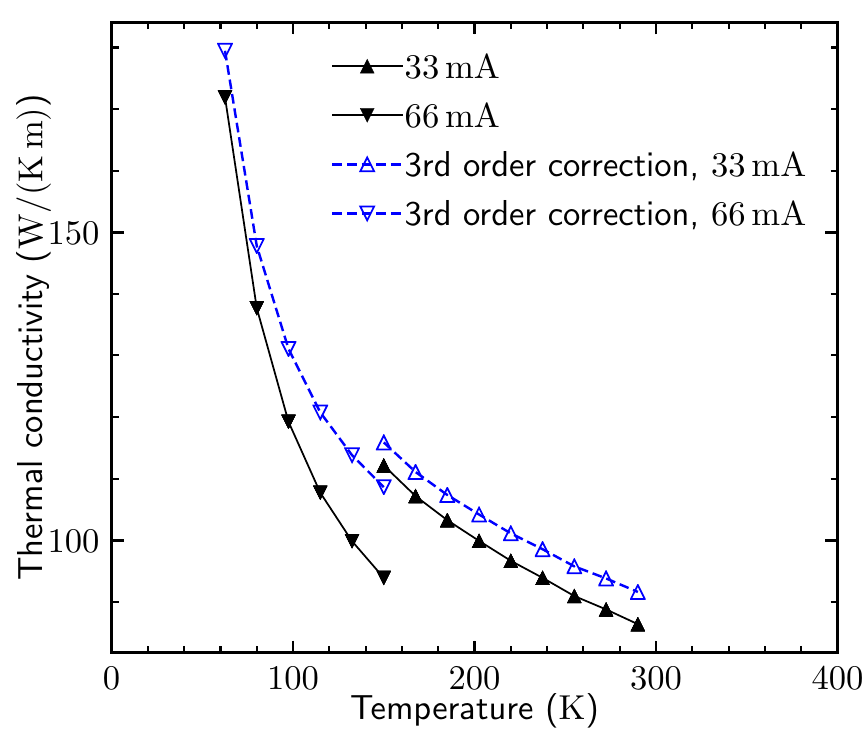}
\caption{Calculated values for $\kappa$ with and without 3rd order correction terms. The higher the current, the more necessary it becomes to take into account correction terms.}
\label{fig: kappa_run3_run4_corrections}
\end{figure}

We note here also that Equation \ref{eqn:V3approx} shows that the $3\omega$ voltage is not expected to depend on $\Delta T$ at all, in line with experimental observations. Equation \ref{eqn:V3approx} can likewise be expanded around $\mu$ to give:
\begin{equation}
\begin{split}
 \lim_{\omega\to 0} V_{3\omega}(I) \approx & \frac{\alpha R^2 I^3}{48 K}
 \biggl( 1 + \frac{\alpha R I^2}{10K}\\ 
 &+ \frac{17\alpha^2 R^2 I^4}{1680 K^2}+ \dots \biggr) + O(\mu^2)
\end{split}
 \label{eqn: correction_thermal_capacitance}
\end{equation}
The leading term in the expansion is the familiar solution of the $3\omega$ thermal conductivity experiment, while the following terms are the corrections that are required to obtain $K$ when the applied current is high.
Fig. \ref{fig: kappa_run3_run4_corrections} shows $\kappa$ values obtained without correction and with a correction using the three leading terms. Although there is improvement, the overlap between ranges is still imperfect. We believe this arises because of the temperature dependence of $K$, which is not accounted for in Eqn.~\ref{eqn: correction_thermal_capacitance}. This is included in our finite element model (Appendix~\ref{sec:finite}).

\subsection{Solution in the limit $J_1\rightarrow 0$}

Consider the heat equation with a sinusoidal source term:
\begin{equation}
\kappa \frac{\partial^2 T}{\partial x^2} + Q_1\sin(\omega_H t) = C_V\frac{\partial T}{\partial t}
\end{equation}
where $Q_1$ is a constant defining the amplitude of the heating, which occurs at frequency $\omega_H$. The solution to this equation with boundary conditions $T(\pm l) = T_0$ has been obtained in series\cite{Lu_lumped_2001} and closed\cite{dames20051} form, and we assume that the solution for boundary conditions $T(\pm l) = T_0 \pm \frac{\Delta T}{2}$ can be constructed by adding $x\Delta T/2l$.

The approximation used to calculate $3\omega$ voltages in Ref.~\onlinecite{dames20051} is that the oscillatory part of the Joule heating is:
\begin{equation}
\begin{split}
\tilde{\rho J^2} &= -(\rho_0+\rho_1 T)\frac{J_1^2}{2}\cos(2\omega t)\\
       &\approx -\rho_0 \frac{J_1^2}{2}\cos(2\omega t)\\
\end{split}
\end{equation}
Hence $Q_1 = \frac{1}{2}\rho_0 J_1^2$ and $\omega_H = 2\omega$. This approximation becomes exact in the limit that the applied current is so low that no heating is caused, i.e. $J_1\rightarrow 0$.

The $3\omega$ voltages are:
\begin{equation}
\begin{split}
X_{3\omega} &= -\frac{\rho_0^2 \alpha J_1^3 l^3}{\kappa q_2^3} \biggl[ \frac{\sinh(q_2)-\sin(q_2)}{\cosh(q_2)+\cos(q_2)} \biggr]\\
Y_{3\omega} &= \frac{\rho_0^2 \alpha J_1^3 l^3}{\kappa q_2^3} \biggl[ \frac{\sinh(q_2)+\sin(q_2)}{\cosh(q_2)+\cos(q_2)} -q_2 \biggr]\\
\end{split}
\end{equation} where $q_2 = 2l\sqrt{\frac{\omega C_V}{\kappa}}$.

We can also approximate the most relevant contribution to the Thomson heating as:
\begin{equation}
\mu J \frac{d T}{dx} \approx \mu J_1 \sin(\omega t) \frac{\Delta T}{2l}
\end{equation}
given the argument outlined in the introduction, that most oscillatory contributions generated by the oscillating gradient $\frac{dT}{dx}$ cancel out over the length of the wire. The heating is assumed to be so low that there is no appreciable deviation from this assumption, hence the same validity in the limit $J_1\rightarrow 0$. In this case, $Q_1 = \mu J_1 \frac{\Delta T}{2l}$ and $\omega_H = \omega$. 

The $2\omega$ voltages arising from the solution are:
\begin{equation}
\begin{split}
X_{2\omega} &= \frac{2 \rho \alpha J_1^2 \mu \frac{\Delta T}{2l} l^3}{\kappa q_1^3} \biggl[ \frac{\sinh(q_1)+\sin(q_1)}{\cosh(q_1)+\cos(q_1)} -q_1 \biggr]\\
Y_{2\omega} &= \frac{2 \rho \alpha J_1^2 \mu \frac{\Delta T}{2l} l^3}{\kappa (2lq_1)^3} \biggl[ \frac{\sinh(q_1)-\sin(q_1)}{\cosh(q_1)+\cos(q_1)} \biggr]
\end{split}
\end{equation} where $q_1 = 2l \sqrt{\frac{\omega C_V}{2\kappa}}$.

In the low frequency limit, we obtain:
\begin{equation}
\mu = -\lim_{\omega \rightarrow 0}\frac{Y_{2\omega}}{X_{3\omega}}\frac{\rho J_1 }{2\frac{\Delta T}{2l}} = -\lim_{\omega \rightarrow 0}\frac{Y_{2\omega}}{X_{3\omega}}\frac{RI}{2 \Delta T}
\label{eqn:mulumped}
\end{equation}
which is the first term of Eqn.~\ref{eqn:approximation} as well as:
\begin{equation}
\kappa = -\lim_{\omega \rightarrow 0}\frac{1}{X_{3\omega}}\frac{\alpha \rho^2 J_1^3 l^3}{6} = -\lim_{\omega \rightarrow 0}\frac{1}{X_{3\omega}}\frac{\alpha R^2 I^3}{48} \frac{2l}{A}
\label{eqn:kappalumped}
\end{equation}
which is the first term of Eqn.~\ref{eqn: correction_thermal_capacitance}.

The above equations indicate that we expect $X_{3\omega} \sim J_1^3$ and $Y_{2\omega} \sim J_1^2$. Because we work in the regime where the current is high enough to produce a significant temperature rise in the wire (e.g. 10K at 300K), these scaling relations are not obeyed either in experimental data or in full simulations. It is necessary to use such high currents to have an observable $Y_{2\omega}$ voltage. 

The deviations have a significant impact on the evaluation of $\kappa$ using Eqn.~\ref{eqn:kappalumped}, rendering the extracted values unreliable. However, it is remarkable that the formula for $\mu$ is, to first order, the same as the more accurate result obtained through series expansion of the exact steady-state solution. This arises because the deviations from pure $J_1^2$ and $J_1^3$ behaviour nearly cancel in the ratio $\frac{Y_{2\omega}}{X_{3\omega}}$. Even though the value of $\kappa$ is unreliable, the value of $\mu$ can be extracted with little error by measuring only two voltages.

\subsection{Full solution for $\mu=0$}
We obtain a full solution for the one-dimensional Joule-only heat equation with constant $\kappa$, $C_V$, $\rho = \rho_0+\rho_1 T$ and $J = J_1\sin(\omega t)$ in the form of a series of convolution integrals. This can be computed numerically much more quickly than the finite element simulations, and is therefore of use in calculating $\kappa$ and $C_V$.

We use an integrating factor $P(t)$ to simplify Equation~\ref{eqn:fullheat}:
\begin{equation}
P(t) = \frac{\rho_1 J_1^2}{2 C_V} \left[t - \frac{\sin (2\omega t)}{2\omega} \right] 
\end{equation}

If we write
\begin{equation}
T(x,t) = \frac{\rho_0}{\rho_1}[e^{P(t)} - 1] + e^{P(t)}T_1(x,t)
\end{equation}
then Equation~\ref{eqn:fullheat} becomes
\begin{equation}
\kappa \frac{\partial^2 T_1}{\partial x^2} -\mu J_1 \sin(\omega t) \frac{\partial T_1}{\partial x} = C_V\frac{\partial T_1}{\partial t}
\end{equation}
with boundary conditions
\begin{equation}
T_1(\pm l,t) = e^{-P(t)} \bigg ( T_0 + \frac{\rho_0}{\rho_1} \pm \frac{\Delta T_0}{2} \bigg ) - \frac{\rho_0}{\rho_1} \forall t
\end{equation}
and initial condition $T_1(x,0) = T_0 + \frac{x\Delta T_0}{2l}$.

While this remains difficult to solve, if we set $\mu=0$ then a solution can be obtained from Ref.~\onlinecite{carslaw1965conduction}, sec.3.5. The solution is the sum of the boundary value $T_1^b(x,t)$ and initial value $T_1^i(x,t)$ problems:
\begin{equation}
\begin{split}
 T_1^b(\pm l,t) &= T_1(\pm l,t), \qquad T_1^b(x,0) = 0 \\
 T_1^i(\pm l,t) &= 0, \qquad \qquad T_1^i(x,0) = T_1(x,0)
\end{split}
\end{equation}

To obtain the solution to the boundary value problem in a slightly simplified form compared to Ref.~\onlinecite{carslaw1965conduction}, we define $Q(t) = e^{-P(t)}-1$ and exploit the following properties of Laplace transforms to recast the solution:
\begin{equation}
 \begin{split}
  \bar{f}(s) = &\Lapl [f(t)] = \int_0^\infty e^{-st} f(t) dt\\
  &\Lapl \left[\frac{\partial f(t)}{\partial t} \right] = [f(t)e^{-st}]^\infty_0 + s\bar{f}(s)\\
  &\Lapl \left[ \int_0^t f_1(t-\tau) f_2(\tau) d\tau \right] = \bar{f_1}(s)\bar{f_2}(s)
 \end{split}
\end{equation}
We also split the solution into even and odd parts with respect to $x$. For the new boundary conditions $\phi^{\text{even}}(t) = \frac{1}{2}[T_1(l,t)+T_1(-l,t)]$ and $\phi^{\text{odd}}(t) = \frac{1}{2}[T_1(l,t)-T_1(-l,t)]$, the general solution over the interval $-l<x<l$ is:
\begin{equation}
\begin{split}
 T_1^{\text{b,even}}(x,t) &= \int_0^t\phi^{\text{even}}(\tau)\frac{\partial F^{\text{even}}}{\partial t}(x,t-\tau)d\tau \\
 T_1^{\text{b,odd}}(x,t) &= \int_0^t\phi^{\text{odd}}(\tau)\frac{\partial F^{\text{odd}}}{\partial t}(x,t-\tau)d\tau \\
 \label{eqn:boundarysolution}
\end{split}
\end{equation}
where
\begin{equation}
 \begin{split}
  F^{\text{even}}(x,t) &= 1 - \frac{4}{\pi}\sum^\infty_{m=0} \frac{(-1)^m}{(2m+1)}\cos\left[ \frac{(2m+1)\pi x}{2l}\right]\\
  & \quad\times \exp \left[-\frac{\kappa}{C_V}\frac{(2m+1)^2\pi^2 t}{4l^2} \right] \\
  F^{\text{odd}}(x,t) &= \frac{x}{l} + \frac{2}{\pi}\sum^\infty_{m=1} \frac{(-1)^m}{m}\cos\left( \frac{m\pi x}{l}\right) e^{ -\frac{\kappa}{C_V}\frac{m^2\pi^2 t}{l^2} } 
 \end{split}
\end{equation}

For our particular problem, performing the forward and inverse Laplace transformations leads to:
\begin{equation}
 \begin{split}
 T_1^{\text{b,even}}(x,t) =& T_0 F^{\text{even}}(x,t)\\
 +& \left( T_0 + \frac{\rho_0}{\rho_1}\right) \int_0^t \frac{\partial Q(\tau)}{\partial \tau} F^{\text{even}}(x,t-\tau) d\tau \\
 T_1^{\text{b,odd}}(x,t) =& \frac{\Delta T_0}{2} F^{\text{odd}}(x,t)\\
  +& \frac{\Delta T_0}{2}\int_0^t\frac{\partial Q(\tau)}{\partial \tau}F^{\text{odd}}(x,t-\tau)d\tau \\
 \end{split}
\end{equation}
These convolution integrals are quick to evaluate numerically (relative to the finite element simulations) as they need only be done once for the entire time range of interest.

We split the initial value problem similarly over the initial conditions $\psi^{\text{even}}(x) = \frac{1}{2}[T_1(x,0)+T_1(-x,0)]$ and $\psi^{\text{odd}}(x) = \frac{1}{2}[T_1(x,0)-T_1(-x,0)]$ so that the generic solution reads:
\begin{equation}
 \begin{split}
  T_1^{\text{i,even}} &= \frac{1}{l}\sum_{m=0}^\infty \cos \left[ \frac{(2m+1)\pi x}{2l} \right] e^{ -\frac{\kappa}{C_V}\frac{(2m+1)^2\pi^2 t}{4l^2} } \\
  &\times\int_{-l}^{l} \psi^{\text{even}}(x^\prime)\cos \left[ \frac{(2m+1)\pi x^\prime}{2l} \right]  dx^\prime\\
  T_1^{\text{i,odd}} &= \frac{1}{l}\sum_{m=1}^\infty \sin\left( \frac{m\pi x}{l} \right) \exp \left[ -\frac{\kappa}{C_V}\frac{m^2\pi^2 t}{l^2} \right] \\
  &\times \int_{-l}^{l} \psi^{\text{odd}}(x^\prime)\sin \left( \frac{m\pi x^\prime}{l} \right)  dx^\prime\\
 \end{split}
\end{equation}
Applying our particular initial condition to obtain the Fourier series coefficients gives:
\begin{equation}
 \begin{split}
  T_1^{\text{i,even}} &= \sum_{m=0}^\infty 2T_0 \frac{(-1)^m}{(2m+1)\pi} \cos \left[ \frac{(2m+1)\pi x}{2l} \right] \\ 
  &\times \exp \left[ -\frac{\kappa}{C_V}\frac{(2m+1)^2\pi^2 t}{4l^2}  \right] \\
  T_1^{\text{i,odd}} &= \sum_{m=1}^\infty \Delta T_0 \frac{(-1)^{(m+1)}}{m \pi} \sin \left( \frac{m\pi x}{l} \right) e^{ -\frac{\kappa}{C_V}\frac{m^2\pi^2 t}{l^2} } \\
 \end{split}
\end{equation}

This solution is of some use in the case of platinum, where the resistivity is nearly linear in temperature in a large temperature range, where $\kappa$ and $C_V$ are coincidentally nearly constant.

\section{Finite Element Analysis}
\label{sec:finite}

Exact solutions to the heat equation are available only in particular simplified cases. We use a 1D finite element model to resolve the heat equation with its associated Dirichlet boundary conditions to take into account: DC offset current; thermal radiation losses; non-linear resistivity; temperature dependent physical parameters such as $\kappa$, $C_V$ and $\mu$. All of these can be easily treated within a numerical approach.

Our finite element simulation is written in Python using FEniCS and employs a simple backward Euler finite difference to model the time-dependence of Eqn.~\ref{eqn:fullheat}. 
The main input is the temperature-dependent resistance of the wire.
The equation used is:
\begin{equation}
\begin{split}
    \nabla &\cdot ([\kappa_0+\kappa_1 (T-T_0)] \nabla T) - \mu [J_0 + J_1\sin(\omega t)] \nabla T \\
    &+\rho(T)[J_0 + J_1\sin(\omega t)]^2 -\frac{4}{d} \varepsilon \sigma_{S-B} (T^4-T_{\text{ref}}^4)\\
    &= C_V \frac{[T - T_{-1}]}{\Delta t}
\end{split}
\end{equation}
where $\kappa = \kappa_0+\kappa_1 (T-T_0)$ models the temperature dependence of the thermal conductivity, $J_0$ is the DC offset current density, $\rho(T)$ is interpolated from the measured resistance of the wire, $d$ is the diameter of the wire, $\varepsilon$ is the emissivity of the wire, $\sigma_{S-B}$ is the Stefan-Boltzmann constant, $T_{-1}$ is the temperature profile from the previous time step and $\Delta t$ is the time step. The boundary conditions are $T(\pm l, t) = T_0 \pm \Delta T/2$.

The results of the simulations have been rigorously compared to the exact solutions presented in Appendix \ref{sec:solutions}, where this temperature dependence takes a simple linear form. They are in perfect agreement. Once the temperature profile is obtained at any one time, it is trivial to extract the voltage between the ends of the wire by numerical integration of $V = \int_{-l}^{l} \rho(T(x)) J dx$.

We find that a spatial resolution of 32 elements along the wire and a temporal resolution of $\bar{t}/128$ where $\bar{t}=\frac{2\pi}{\omega}$ is enough to ensure convergence of the simulations to within \SIrange[range-phrase = --]{1}{3}{\percent}, lower than the noise level in the $2\omega$ data.

At each temperature, we adjust the input parameters $\kappa$, $\mu$ and $C_V$ to obtain the best least-squares fit to the experimental data. In practice, since the simulated $3\omega$ voltage depends almost exclusively on $\kappa$ and $C_V$, these are adjusted in a first step. Then the value of $\mu$ is adjusted while keeping $\kappa$ and $C_V$ fixed to obtain the correct $2\omega$ voltages. The fitting was performed computationally using a non-linear least-squares fitting routine (specifically the Nelder-Mead algorithm from the Scipy python package `minimize'). The convergence criterion was chosen such that the free parameters had converged to within \num{1}\% (for $\kappa$ and $C_V$) and at worst \num{3}\% for $\mu$. 

We note that the precise dimensions of the wire are not required to produce realistic values of $\kappa$ and $\mu$, as long as the combination of simulated cross-section, wire length and resistivity reproduce the experimental resistance at each temperature. This leads to the lack of geometric factors in Eqns.~\ref{eqn:approximation} and \ref{eqn: correction_thermal_capacitance}.

\subsection{DC offset current}
It is simple to include a finite DC offset current in the simulation such that $J = J_0 + J_1\sin\omega t$. This models the real-world current source, where such an offset is expected. The simulations confirm our observations and expectations. There is no effect on the $3\omega$ signal for reasonable values such that $J_0 \sim 10^{-3}J_1$. There is an additional contribution to the $2\omega$ voltage, but this does not depend on the value of $\Delta T$ and can therefore be excluded by subtracting a baseline value at each frequency.

\subsection{Radiative Losses}
\label{subsec: radlosses} 
We add a standard radiative term $(4/d)\varepsilon \sigma_\text{S-B}(T^4-T_\text{ref}^4)$ to the 1D simulation, scaled appropriately to reflect the real dimensions of the wire. $\varepsilon = 0.1$ is an appropriate value for the emissivity for both Pt and Ni wires over a large temperature range. We treat the radiation term as a volumetric source instead of a surface flux by scaling it using the factor $4/d$, where $d$ is the diameter of the wire. This is appropriate when the heat equation is being resolved in 1D, but ignores end effects.

We perform these simulations at a number of elevated temperatures for the Pt wire, extrapolating the resistance linearly from our data, but keeping the input values of $\kappa$ and $\mu$ constant. We then fit the simulated data \textit{without} the radiative term to obtain the values of $\kappa$ and $\mu$ that would be observed if radiation were neglected. 

\begin{figure}
\includegraphics{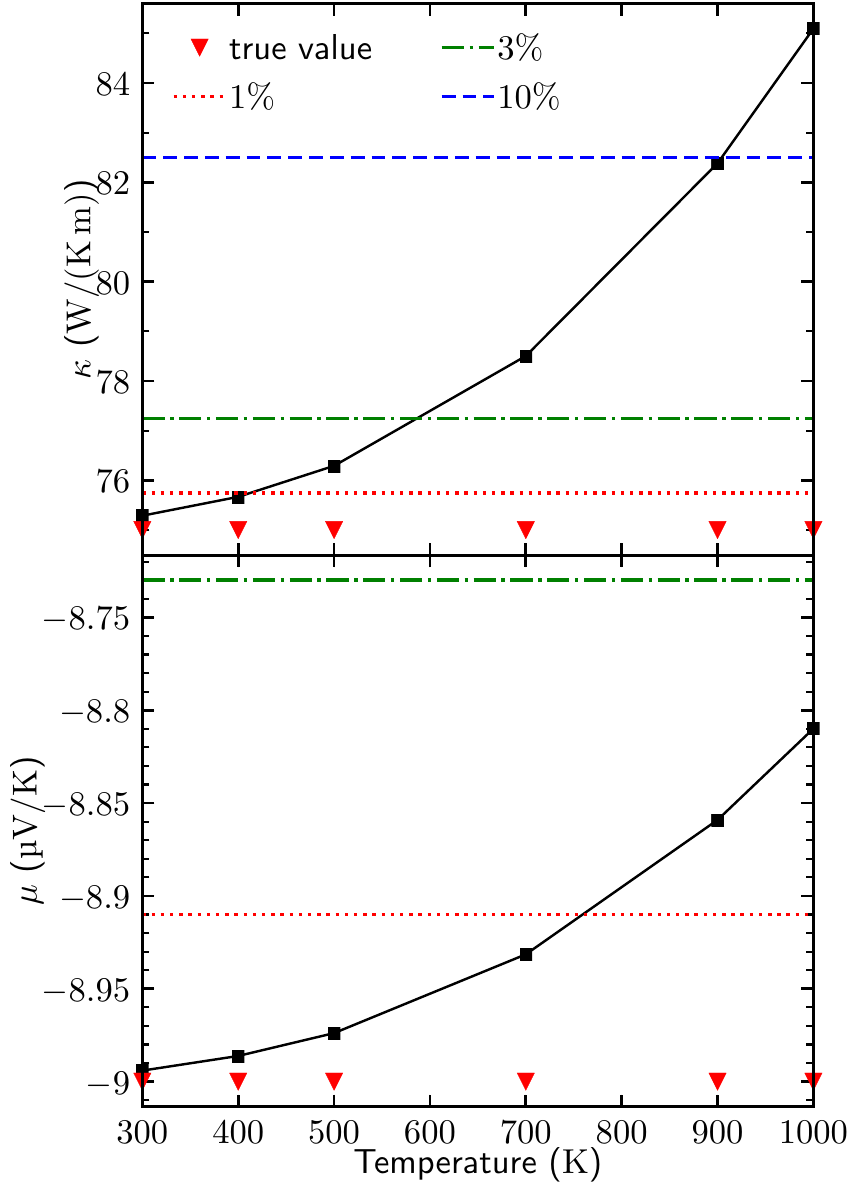}
\caption{The apparent values of $\kappa$ and $\mu$ that one would extract if one were to neglect radiation effects are shown in black squares. The true values of $\kappa$ and $\mu$ are kept constant throughout the whole temperature range (red triangles). The \SIlist{1;3;10}{\percent} thresholds for errors in the apparent values relative to the true ones are shown as horizontal dashed lines.}
\label{fig:rad_plot}
\end{figure}

The results are shown in Fig.~\ref{fig:rad_plot}. The effect of neglecting radiative loss is that $\kappa$ is systematically overestimated. The error is $\SI{1}{\percent}$ at \SI{410}{\kelvin}, $\SI{3}{\percent}$ at \SI{590}{\kelvin}, and $\SI{10}{\percent}$ at $\SI{905}{\kelvin}$. The magnitude of $\mu$ is underestimated, but the effect is much smaller, $\SI{1}{\percent}$ at \SI{750}{\kelvin} and $\SI{3}{\percent}$ at a significantly higher temperature than \SI{1000}{\kelvin}. It is worth mentioning that $C_V$ is even less affected, its apparent value is smaller by less than $\SI{1}{\percent}$ at temperatures as high as \SI{1000}{\kelvin}.

\subsection{Non-linear resistivity}
The temperature dependence of the resistance of the wire is incorporated into the finite element model using a look-up table. This allows us to go beyond the exact solutions of the heat equation. It is particularly important in the case of nickel, where the resistance is strongly non-linear with temperature, but less so for platinum.

\subsection{Temperature dependence of $\kappa$}
We found that in order to fit the data more precisely, it was necessary in some temperature ranges to include the temperature dependence of $\kappa$ in the simulation, such that $\kappa(T) = \kappa_0 + \kappa_1 (T-T_0)$. Introducing this extra parameter causes the fitting for a single value of $J$ to be over-determined. We therefore systematically applied three different currents to the wire at each reference temperature $T_0$, in order to determine $\kappa_0$ and $\kappa_1$. 

\section{DC measurement by current reversal}
\label{sec:dc}
In principle, we could take DC measurements of the voltages $V(I)$ and $V(-I)$ and use Equation \ref{eqn:nondimvoltage} to evaluate $\mu$ and $\kappa$. This is subject to several practical difficulties. The typical order of magnitude of $V(I)$ is perhaps $\SI{10}{\milli\volt}$. The Thomson signal is of the order of $\SI{1}{\micro\volt}$, meaning that $V(I)+V(-I)\sim \SI{1}{\micro\volt}$. To determine this with a precision of $\SI{1}{\percent}$, a resolution of $\SI{10}{\nano\volt}$ or $\SI{1}{\ppm}$ is required. This is already challenging to accomplish with a DC technique. 

In reality, the voltmeter output $V(0)$ is non-zero and drifts over time. Therefore $V(I)+V(-I)\sim \SI{1}{\micro\volt} +\num{2}V(\num{0})$. While $V(0)$ can also be measured, the time in between measurements with and without current is necessarily long to allow complete thermal relaxation. Maintaining $\SI{1}{\ppm}$ stability over these time scales is difficult. A purely DC technique exploiting the resistance of the wire as a thermometer is therefore hard to implement.

\section{Comparison with Amagai \emph{et al}.}
\label{sec:amagai}
In Reference \onlinecite{amagai2015ac}, the authors suspend a \SI{0.5}{\milli\meter} diameter platinum wire between two posts and pass currents up to \SI{2.5}{\ampere} through it. They measure the temperature rise along the wire using type K fine-wire thermocouples. They develop a thermal model to account for heat loss by the thermocouple induced by the measurement. Joule heating causes a temperature rise of approximately \SI{4}{\kelvin}, while Thomson heating generates an additional $\pm$\SI{40}{\milli\kelvin} at most, depending on the sign of the current and location of the thermocouple. This additional temperature change is directly proportional to the Thomson coefficient. The stated voltmeter resolution is \SI{100}{\nano\volt}, which corresponds to a temperature resolution of $\pm\SI{10}{\milli\kelvin}$, or an uncertainty in the Thomson-related temperature difference of $\SI{25}{\percent}$. Several thermocouple readings along the length of the wire are therefore required to reduce this uncertainty to an acceptable level.

Although the authors of Ref. \onlinecite{amagai2015ac} use a high-frequency AC current to establish a Joule-only benchmark reading, the temperature differences are still collected by measuring the thermocouple response using a DC voltmeter. This is therefore susceptible to the same kinds of problems discussed in Appendix \ref{sec:dc}. Maintaining a stable reading over several hundred seconds to better than \SI{100}{\nano\volt} is a challenge. A phase-sensitive alternating current technique does not suffer from these drawbacks. The thermocouple technique also requires that the sample wire is physically contacted in the middle, which introduces further uncertainty and complexity to the measurement.

\section*{References}

\nocite{*} 
\bibliography{thomson_effect}

\begin{thebibliography}{19}%
\makeatletter
\providecommand \@ifxundefined [1]{%
 \@ifx{#1\undefined}
}%
\providecommand \@ifnum [1]{%
 \ifnum #1\expandafter \@firstoftwo
 \else \expandafter \@secondoftwo
 \fi
}%
\providecommand \@ifx [1]{%
 \ifx #1\expandafter \@firstoftwo
 \else \expandafter \@secondoftwo
 \fi
}%
\providecommand \natexlab [1]{#1}%
\providecommand \enquote  [1]{``#1''}%
\providecommand \bibnamefont  [1]{#1}%
\providecommand \bibfnamefont [1]{#1}%
\providecommand \citenamefont [1]{#1}%
\providecommand \href@noop [0]{\@secondoftwo}%
\providecommand \href [0]{\begingroup \@sanitize@url \@href}%
\providecommand \@href[1]{\@@startlink{#1}\@@href}%
\providecommand \@@href[1]{\endgroup#1\@@endlink}%
\providecommand \@sanitize@url [0]{\catcode `\\12\catcode `\$12\catcode
  `\&12\catcode `\#12\catcode `\^12\catcode `\_12\catcode `\%12\relax}%
\providecommand \@@startlink[1]{}%
\providecommand \@@endlink[0]{}%
\providecommand \url  [0]{\begingroup\@sanitize@url \@url }%
\providecommand \@url [1]{\endgroup\@href {#1}{\urlprefix }}%
\providecommand \urlprefix  [0]{URL }%
\providecommand \Eprint [0]{\href }%
\providecommand \doibase [0]{http://dx.doi.org/}%
\providecommand \selectlanguage [0]{\@gobble}%
\providecommand \bibinfo  [0]{\@secondoftwo}%
\providecommand \bibfield  [0]{\@secondoftwo}%
\providecommand \translation [1]{[#1]}%
\providecommand \BibitemOpen [0]{}%
\providecommand \bibitemStop [0]{}%
\providecommand \bibitemNoStop [0]{.\EOS\space}%
\providecommand \EOS [0]{\spacefactor3000\relax}%
\providecommand \BibitemShut  [1]{\csname bibitem#1\endcsname}%
\let\auto@bib@innerbib\@empty
\bibitem [{\citenamefont {Thomson}(1857)}]{thomson18574}%
  \BibitemOpen
  \bibfield  {author} {\bibinfo {author} {\bibfnamefont {W.}~\bibnamefont
  {Thomson}},\ }\href@noop {} {\bibfield  {journal} {\bibinfo  {journal}
  {Proceedings of the Royal Society of Edinburgh}\ }\textbf {\bibinfo {volume}
  {3}},\ \bibinfo {pages} {91} (\bibinfo {year} {1857})}\BibitemShut {NoStop}%
\bibitem [{\citenamefont {Thomson}(1856)}]{thomson1853}%
  \BibitemOpen
  \bibfield  {author} {\bibinfo {author} {\bibfnamefont {W.}~\bibnamefont
  {Thomson}},\ }\href {http://www.jstor.org/stable/41206195} {\bibfield
  {journal} {\bibinfo  {journal} {Philosophical Transactions of the Royal
  Society of London}\ }\textbf {\bibinfo {volume} {146}},\ \bibinfo {pages}
  {649} (\bibinfo {year} {1856})}\BibitemShut {NoStop}%
\bibitem [{\citenamefont {Onsager}(1931)}]{onsager1931}%
  \BibitemOpen
  \bibfield  {author} {\bibinfo {author} {\bibfnamefont {L.}~\bibnamefont
  {Onsager}},\ }\href {\doibase 10.1103/PhysRev.37.405} {\bibfield  {journal}
  {\bibinfo  {journal} {Phys. Rev.}\ }\textbf {\bibinfo {volume} {37}},\
  \bibinfo {pages} {405} (\bibinfo {year} {1931})}\BibitemShut {NoStop}%
\bibitem [{\citenamefont {Chen}, \citenamefont {Yan},\ and\ \citenamefont
  {Wu}(1996)}]{teg-most-important}%
  \BibitemOpen
  \bibfield  {author} {\bibinfo {author} {\bibfnamefont {J.}~\bibnamefont
  {Chen}}, \bibinfo {author} {\bibfnamefont {Z.}~\bibnamefont {Yan}}, \ and\
  \bibinfo {author} {\bibfnamefont {L.}~\bibnamefont {Wu}},\ }\href {\doibase
  10.1063/1.362507} {\bibfield  {journal} {\bibinfo  {journal} {Journal of
  Applied Physics}\ }\textbf {\bibinfo {volume} {79}},\ \bibinfo {pages} {8823}
  (\bibinfo {year} {1996})}\BibitemShut {NoStop}%
\bibitem [{\citenamefont {Borelius}, \citenamefont {Keesom},\ and\
  \citenamefont {H.}(1928)}]{borelius1928}%
  \BibitemOpen
  \bibfield  {author} {\bibinfo {author} {\bibfnamefont {G.}~\bibnamefont
  {Borelius}}, \bibinfo {author} {\bibfnamefont {W.~H.}\ \bibnamefont
  {Keesom}}, \ and\ \bibinfo {author} {\bibfnamefont {J.~C.}\ \bibnamefont
  {H.}},\ }\href
  {http://www.dwc.knaw.nl/toegangen/digital-library-knaw/?pagetype=publDetail&pId=PU00015666&lang=en}
  {\bibfield  {journal} {\bibinfo  {journal} {Proc. KNAW}\ }\textbf {\bibinfo
  {volume} {31}},\ \bibinfo {pages} {1046} (\bibinfo {year}
  {1928})}\BibitemShut {NoStop}%
\bibitem [{\citenamefont {Lander}(1948)}]{lander1948}%
  \BibitemOpen
  \bibfield  {author} {\bibinfo {author} {\bibfnamefont {J.~J.}\ \bibnamefont
  {Lander}},\ }\href {\doibase 10.1103/PhysRev.74.479} {\bibfield  {journal}
  {\bibinfo  {journal} {Phys. Rev.}\ }\textbf {\bibinfo {volume} {74}},\
  \bibinfo {pages} {479} (\bibinfo {year} {1948})}\BibitemShut {NoStop}%
\bibitem [{\citenamefont {Nystr\"{o}m}(1947)}]{nystrom1947}%
  \BibitemOpen
  \bibfield  {author} {\bibinfo {author} {\bibfnamefont {J.}~\bibnamefont
  {Nystr\"{o}m}},\ }\href@noop {} {\bibfield  {journal} {\bibinfo  {journal}
  {Arkiv För. Matematik, Astronomi och Fysik}\ }\textbf {\bibinfo {volume}
  {34a}},\ \bibinfo {pages} {1} (\bibinfo {year} {1947})}\BibitemShut {NoStop}%
\bibitem [{\citenamefont {Cusack}\ and\ \citenamefont
  {Kendall}(1958)}]{cusack1958}%
  \BibitemOpen
  \bibfield  {author} {\bibinfo {author} {\bibfnamefont {N.}~\bibnamefont
  {Cusack}}\ and\ \bibinfo {author} {\bibfnamefont {P.}~\bibnamefont
  {Kendall}},\ }\href {http://stacks.iop.org/0370-1328/72/i=5/a=429} {\bibfield
   {journal} {\bibinfo  {journal} {Proceedings of the Physical Society}\
  }\textbf {\bibinfo {volume} {72}},\ \bibinfo {pages} {898} (\bibinfo {year}
  {1958})}\BibitemShut {NoStop}%
\bibitem [{\citenamefont {Roberts}(1977)}]{roberts1977absolute}%
  \BibitemOpen
  \bibfield  {author} {\bibinfo {author} {\bibfnamefont {R.}~\bibnamefont
  {Roberts}},\ }\href {https://doi.org/10.1080/00318087708244450} {\bibfield
  {journal} {\bibinfo  {journal} {Philosophical Magazine}\ }\textbf {\bibinfo
  {volume} {36}},\ \bibinfo {pages} {91} (\bibinfo {year} {1977})}\BibitemShut
  {NoStop}%
\bibitem [{\citenamefont {Cahill}(1990)}]{Cahill1990}%
  \BibitemOpen
  \bibfield  {author} {\bibinfo {author} {\bibfnamefont {D.~G.}\ \bibnamefont
  {Cahill}},\ }\href {\doibase 10.1063/1.1141498} {\bibfield  {journal}
  {\bibinfo  {journal} {Review of Scientific Instruments}\ }\textbf {\bibinfo
  {volume} {61}},\ \bibinfo {pages} {802} (\bibinfo {year} {1990})}\BibitemShut
  {NoStop}%
\bibitem [{\citenamefont {Dames}\ and\ \citenamefont
  {Chen}(2005)}]{dames20051}%
  \BibitemOpen
  \bibfield  {author} {\bibinfo {author} {\bibfnamefont {C.}~\bibnamefont
  {Dames}}\ and\ \bibinfo {author} {\bibfnamefont {G.}~\bibnamefont {Chen}},\
  }\href {\doibase 10.1063/1.2130718} {\bibfield  {journal} {\bibinfo
  {journal} {Review of Scientific Instruments}\ }\textbf {\bibinfo {volume}
  {76}},\ \bibinfo {pages} {124902} (\bibinfo {year} {2005})}\BibitemShut
  {NoStop}%
\bibitem [{\citenamefont {Amagai}\ \emph {et~al.}(2015)\citenamefont {Amagai},
  \citenamefont {Yamamoto}, \citenamefont {Akoshima}, \citenamefont {Fujiki},\
  and\ \citenamefont {Kaneko}}]{amagai2015ac}%
  \BibitemOpen
  \bibfield  {author} {\bibinfo {author} {\bibfnamefont {Y.}~\bibnamefont
  {Amagai}}, \bibinfo {author} {\bibfnamefont {A.}~\bibnamefont {Yamamoto}},
  \bibinfo {author} {\bibfnamefont {M.}~\bibnamefont {Akoshima}}, \bibinfo
  {author} {\bibfnamefont {H.}~\bibnamefont {Fujiki}}, \ and\ \bibinfo {author}
  {\bibfnamefont {N.-h.}\ \bibnamefont {Kaneko}},\ }\href {\doibase
  10.1109/TIM.2014.2381752} {\bibfield  {journal} {\bibinfo  {journal} {IEEE
  Transactions on Instrumentation and Measurement}\ }\textbf {\bibinfo {volume}
  {64}},\ \bibinfo {pages} {1576} (\bibinfo {year} {2015})}\BibitemShut
  {NoStop}%
\bibitem [{\citenamefont {Chiang}(1974)}]{Chiang1974}%
  \BibitemOpen
  \bibfield  {author} {\bibinfo {author} {\bibfnamefont {C.~K.}\ \bibnamefont
  {Chiang}},\ }\href {\doibase 10.1063/1.1686803} {\bibfield  {journal}
  {\bibinfo  {journal} {Review of Scientific Instruments}\ }\textbf {\bibinfo
  {volume} {45}},\ \bibinfo {pages} {985} (\bibinfo {year} {1974})}\BibitemShut
  {NoStop}%
\bibitem [{\citenamefont {Chaussy}, \citenamefont {Guessous},\ and\
  \citenamefont {Mazuer}(1981)}]{Chaussy1981}%
  \BibitemOpen
  \bibfield  {author} {\bibinfo {author} {\bibfnamefont {J.}~\bibnamefont
  {Chaussy}}, \bibinfo {author} {\bibfnamefont {A.}~\bibnamefont {Guessous}}, \
  and\ \bibinfo {author} {\bibfnamefont {J.}~\bibnamefont {Mazuer}},\ }\href
  {\doibase 10.1063/1.1136520} {\bibfield  {journal} {\bibinfo  {journal}
  {Review of Scientific Instruments}\ }\textbf {\bibinfo {volume} {52}},\
  \bibinfo {pages} {1721} (\bibinfo {year} {1981})}\BibitemShut {NoStop}%
\bibitem [{\citenamefont {Alnæs}\ \emph {et~al.}(2015)\citenamefont {Alnæs},
  \citenamefont {Blechta}, \citenamefont {Hake}, \citenamefont {Johansson},
  \citenamefont {Kehlet}, \citenamefont {Logg}, \citenamefont {Richardson},
  \citenamefont {Ring}, \citenamefont {Rognes},\ and\ \citenamefont
  {Wells}}]{ans20553}%
  \BibitemOpen
  \bibfield  {author} {\bibinfo {author} {\bibfnamefont {M.}~\bibnamefont
  {Alnæs}}, \bibinfo {author} {\bibfnamefont {J.}~\bibnamefont {Blechta}},
  \bibinfo {author} {\bibfnamefont {J.}~\bibnamefont {Hake}}, \bibinfo {author}
  {\bibfnamefont {A.}~\bibnamefont {Johansson}}, \bibinfo {author}
  {\bibfnamefont {B.}~\bibnamefont {Kehlet}}, \bibinfo {author} {\bibfnamefont
  {A.}~\bibnamefont {Logg}}, \bibinfo {author} {\bibfnamefont {C.}~\bibnamefont
  {Richardson}}, \bibinfo {author} {\bibfnamefont {J.}~\bibnamefont {Ring}},
  \bibinfo {author} {\bibfnamefont {M.}~\bibnamefont {Rognes}}, \ and\ \bibinfo
  {author} {\bibfnamefont {G.}~\bibnamefont {Wells}},\ }\href {\doibase
  10.11588/ans.2015.100.20553} {\bibfield  {journal} {\bibinfo  {journal}
  {Archive of Numerical Software}\ }\textbf {\bibinfo {volume} {3}} (\bibinfo
  {year} {2015}),\ 10.11588/ans.2015.100.20553}\BibitemShut {NoStop}%
\bibitem [{\citenamefont {Lide}(2003)}]{crchandbook}%
  \BibitemOpen
  \bibfield  {author} {\bibinfo {author} {\bibfnamefont {D.}~\bibnamefont
  {Lide}},\ }\href {https://www.crcpress.com} {\emph {\bibinfo {title} {CRC
  Handbook of Chemistry and Physics, 84th Edition}}},\ CRC HANDBOOK OF
  CHEMISTRY AND PHYSICS\ (\bibinfo  {publisher} {Taylor \& Francis},\ \bibinfo
  {year} {2003})\BibitemShut {NoStop}%
\bibitem [{\citenamefont {Roberts}(1981)}]{roberts1981}%
  \BibitemOpen
  \bibfield  {author} {\bibinfo {author} {\bibfnamefont {R.}~\bibnamefont
  {Roberts}},\ }\href {\doibase 10.1080/01418638108222579} {\bibfield
  {journal} {\bibinfo  {journal} {Philosophical Magazine B}\ }\textbf {\bibinfo
  {volume} {43}},\ \bibinfo {pages} {1125} (\bibinfo {year}
  {1981})}\BibitemShut {NoStop}%
\bibitem [{\citenamefont {Lu}, \citenamefont {Yi},\ and\ \citenamefont
  {Zhang}(2001)}]{Lu_lumped_2001}%
  \BibitemOpen
  \bibfield  {author} {\bibinfo {author} {\bibfnamefont {L.}~\bibnamefont
  {Lu}}, \bibinfo {author} {\bibfnamefont {W.}~\bibnamefont {Yi}}, \ and\
  \bibinfo {author} {\bibfnamefont {D.~L.}\ \bibnamefont {Zhang}},\ }\href
  {\doibase 10.1063/1.1378340} {\bibfield  {journal} {\bibinfo  {journal}
  {Review of Scientific Instruments}\ }\textbf {\bibinfo {volume} {72}},\
  \bibinfo {pages} {2996} (\bibinfo {year} {2001})}\BibitemShut {NoStop}%
\bibitem [{\citenamefont {Carslaw}\ and\ \citenamefont
  {Jaeger}(1965)}]{carslaw1965conduction}%
  \BibitemOpen
  \bibfield  {author} {\bibinfo {author} {\bibfnamefont {H.}~\bibnamefont
  {Carslaw}}\ and\ \bibinfo {author} {\bibfnamefont {J.}~\bibnamefont
  {Jaeger}},\ }\href
  {https://global.oup.com/academic/product/conduction-of-heat-in-solids-9780198533689}
  {\emph {\bibinfo {title} {Conduction of Heat in Solids}}}\ (\bibinfo
  {publisher} {Oxford University Press},\ \bibinfo {year} {1965})\BibitemShut
  {NoStop}%
\end{thebibliography}%
\end{document}